\documentclass[aps,tightenlines,superscriptaddress,nofootinbib,showpacs]{revtex4}
\usepackage{amsmath, amsthm, amscd, amssymb, mathtools}
\usepackage{feyn}

 \makeatletter
\newenvironment{sqcases}{%
  \matrix@check\sqcases\env@sqcases
}{%
  \endarray\right.%
}
\def\env@sqcases{%
  \let\@ifnextchar\new@ifnextchar
  \left\lbrack
  \def\arraystretch{1.2}%
  \array{@{}l@{\quad}l@{}}%
}

\begin{document}

\title{
Constraints on the noise in dynamical reduction models}

\author{Kyrylo Simonov}
\email{Kyrylo.Simonov@univie.ac.at}
\affiliation{Faculty of Physics, University of Vienna, Boltzmanngasse 5, 1090 Vienna, Austria}
\author{Beatrix C. Hiesmayr}
\email{Beatrix.Hiesmayr@univie.ac.at}
\affiliation{Faculty of Physics, University of Vienna, Boltzmanngasse 5, 1090 Vienna, Austria}


\begin{abstract} The dynamics of a quantum system with internal degrees of freedom undergoing spontaneous collapse in the position basis are analysed; e.g., neutral mesons or neutrinos. Surprisingly, the value of the Heaviside function $\theta(x)$ at $x=0$ that can in general be chosen in the interval $[0,1]$ leads to different physical predictions.  For the QMUPL (Quantum Mechanics with Universal Position Localization) model only a single value leads to probabilities conserving the particle number.  Herewith the physical properties of the noise field can be constrained. This opens a road to study the physical properties of the noise field essential for collapse models.
\end{abstract}

\maketitle

Dynamical reduction models have been shown to be proper frameworks to circumvent the measurement problem of standard quantum mechanics and allow for a clear definition of a microscopic and a macroscopic system, respectively. In their seminal work~\cite{GRW} Ghirardi, Rimini and Weber introduced a concise framework for microscopic and macroscopic physical systems by introducing a spontaneous collapse of the wave function. Mathematically, this can be achieved by adding specific nonlinear and stochastic terms to the Schrödinger equation. The stochasticity is required to avoid superluminal signaling in general. Up to date such modifications of the unitary evolution have not been found to be in conflict to any experiment. The testability of collapse models is one of its merits and experiments such as, e.g., the ones with $X$-rays~\cite{Catalina1,Catalina2} are putting upper limits on the reduction rate parameter. In this letter we are not interested per se to single out an experimental testable observable, we focus on conceptual issues when considering systems at high energies. We will prove that the mathematical structure of the interaction with the noise field leads to constraints if one imposes particle number conservation.

In general, the dynamics is described by the following non-linear stochastic differential equation~\cite{BassiDuerrHinrichs2011}
\begin{equation}\label{StateVectorEquation}
d|\phi_{t}\rangle=\left[-i\hat{H} dt+\sqrt{\lambda}\sum\limits_{i=1}^N(\hat{A}_i - \langle \hat{A}_i \rangle_t) dW_{i, t}-\frac{\lambda}{2} \sum\limits_{i=1}^N(\hat{A}_i- \langle \hat{A}_i \rangle_t)^2dt\right]|\phi_{t}\rangle,
\end{equation}
with $\hbar=1$ and $\langle \hat{A}_i \rangle_t := \langle \phi_t | \hat{A}_i |\phi_{t}\rangle$. Here $\hat{H}$ is the standard Hamiltonian of the quantum system, $\hat{A}_i$ are a set of $N$ operators introducing the collapse in a certain basis choice (position basis in most cases). $W_{i,t}$ present a set of independent standard Wiener processes and $\lambda$ quantifies the strength of the collapse.

Finding solutions to a non-linear stochastic equation is a cumbersome problem. A first step to simplify the issue is to replace the real noises introduced via $W_{i,t}$ by imaginary ones $i W_{i,t}$. As shown in Ref.~\cite{AdlerBassi2007} this does not change the master equation of the corresponding density operator: the physics is left invariant. Via that ``imaginary transformation'', equation~(\ref{StateVectorEquation}) becomes a Schr\"{o}dinger-like equation with a random Hamiltonian
\begin{equation}
i \frac{d}{dt} |\phi_{t}\rangle = \Bigl[ \hat{H} - \sqrt{\lambda} \sum\limits_{i=1}^N \hat{A}_i w_{i, t} \Bigr] |\phi_{t}\rangle := \Bigl[ \hat{H} + \hat{N}(t) \Bigr] |\phi_{t}\rangle\;,
\end{equation}
where $w_{i,t}:=\frac{dW_{i,t}}{dt}$.

In the following we focus on the QMUPL (Quantum Mechanics with Universal Position Localization) model~\cite{Diosi1989, Diosi1990} for one particle ($N=1$), which is one of the conceptually simplest spontaneous reduction models, however, the results we draw are not limited to this model. Accordingly the operator $\hat{A}$ which induces the collapse is chosen to be the position operator $\hat{q}$ of the particle under interest. We consider $w_{i,t}:=\frac{dW_{i,t}}{dt}$ as a white (uncolored) noise field, where the corresponding correlation function is $\mathbb{E}[w_{i,t} w_{j,s}] = \delta_{ij}\delta(t-s)$. Note that the following results do not depend on the dimension of the position space, therefore for sake of simplicity we stick to the one--dimensional case.

In physics, in particular in high energy physics, there are systems for which the temporal dynamics depends on an additional (internal) degree of freedom. Such systems are for example neutral mesons that show a particle-antiparticle oscillation or flavor-oscillating neutrinos. Without loss of generality we examine neutral K-mesons that are experimentally intensively studied. These neutral kaons interact strongly and decay via the weak interaction. Since both the particle state $|K^0\rangle$ and the antiparticle state $|\bar K^0\rangle$ can decay into the same final states, neutral kaons have to be considered as a two-state system. Moreover, diagonalising the phenomenological Hamiltonian leads to two different mass eigenstates, the short-lived state $|K_S\rangle$ and the long-lived state $|K_L\rangle$, which have distinct masses $m_S,m_L$ and considerably different decay rates $\Gamma_S\approx 600\Gamma_L$. For the sake of simplicity we neglect the small violation of the charge-conjugation-parity symmetry leading to $\langle K_L|K_S\rangle\not= 0$.

To describe the collapse dynamics in the case of neutral kaons we extend the collapse operator $\hat{A}$ of the QMUPL model by also including an operator which acts in flavor space
\begin{eqnarray}
\hat{A} = \hat{q} \otimes \Bigl[ \frac{m_L}{m_0} | K_L \rangle \langle K_L | + \frac{m_S}{m_0} | K_S \rangle \langle K_S | \Bigr],
\end{eqnarray}
where $m_0$ is a reference mass which is taken usually to be the nucleon mass.

We are interested in computing the probabilities in the case of when a certain mass eigenstate $|K_S\rangle,|K_L \rangle$ or strangeness eigenstate $|K^0\rangle, |\bar K^0\rangle$ is produced and one of these states is found at a later time point $t$ in a final state, i.e.
\begin{align}
& P_{K_\mu \rightarrow K_\nu}(p_i, \alpha ; t) = \sum_{p_f} \mathbb{E} | \langle K_\nu, p_f | K_\mu (t), p_i, \alpha \rangle |^2,
\end{align}
where $\mathbb{E}$ denotes the noise average. In order to keep the computations simple we assume that the initial state is a wave packet with width $\sqrt{\alpha}$ in position space and momentum $p_i$ and the final state is a momentum eigenstate.

In order to derive the transition amplitudes we move to the interaction picture. In this case the states evolve due to the noise term $\hat{N}(t)$ in the Schr\"{o}dinger equation and the transition amplitudes in the mass-eigenstate basis are given by
\begin{align}
& T_{\mu\nu}(p_f, p_i, \alpha ; t) := \langle K_{\nu}, p_f | K_{\mu} (t), p_i, \alpha \rangle = e^{-i m_{\mu} t}\; \langle K_{\nu}, p_f | \hat{U}_I(t) | K_{\mu}, p_i, \alpha \rangle,
\end{align}
where $m_{\mu}$ is the mass of one of the lifetime states ($\mu=S,L$). The evolution operator $\hat{U}_I(t)$ is taken in the interaction picture. Since there is no way to solve the differential equations directly, we treat the noise term $\hat{N}(t)$ as a perturbation and expand the evolution operator into the Dyson series. Therefore, up to fourth order, transition amplitudes become
\begin{align}
& \nonumber T_{\mu \nu}(t) \simeq e^{-i m_{\mu} t} \Bigl( T^{(0)}_{\mu\nu}(p_f, p_i, \alpha ; t) + T^{(1)}_{\mu\nu}(p_f, p_i, \alpha ; t) + T^{(2)}_{\mu\nu}(p_f, p_i, \alpha ; t) + T^{(3)}_{\mu\nu}(p_f, p_i, \alpha ; t) + T^{(4)}_{\mu\nu}(p_f, p_i, \alpha ; t)\Bigr),
\end{align}
with
\begin{align}
 & T^{(0)}_{\mu\nu}(p_f, p_i, \alpha ; t) = \langle K_{\nu}, p_f | K_{\mu}, p_i, \alpha \rangle, \\
 & T^{(1)}_{\mu\nu}(p_f, p_i, \alpha ; t) = -i \int\limits_0^t dt_1 \langle K_{\nu}, p_f | \hat{N}_I (t_1) | K_{\mu}, p_i, \alpha \rangle, \\
 & T^{(2)}_{\mu\nu}(p_f, p_i, \alpha ; t) = -\int\limits_0^t dt_1\int\limits_0^{t_1} dt_2 \langle K_{\nu}, p_f | \hat{N}_I (t_1) \hat{N}_I (t_2) | K_{\mu}, p_i, \alpha \rangle, \\
 & T^{(3)}_{\mu\nu}(p_f, p_i, \alpha ; t) = i\int\limits_0^t dt_1\int\limits_0^{t_1} dt_2\int\limits_0^{t_2} dt_3 \langle K_{\nu}, p_f | \hat{N}_I (t_1) \hat{N}_I (t_2) \hat{N}_I (t_3) | K_{\mu}, p_i, \alpha \rangle, \\
 & T^{(4)}_{\mu\nu}(p_f, p_i, \alpha ; t) = \int\limits_0^t dt_1\int\limits_0^{t_1} dt_2\int\limits_0^{t_2} dt_3\int\limits_0^{t_3} dt_4 \langle K_{\nu}, p_f | \hat{N}_I (t_1) \hat{N}_I (t_2) \hat{N}_I (t_3) \hat{N}_I (t_4) | K_{\mu}, p_i, \alpha \rangle,
\end{align}
where $\hat{N}_I(t)$ is the noise term in the interaction picture. The transition probabilities are then given by keeping terms up to second order
\begin{align}
\nonumber P_{K_{\mu}\rightarrow K_{\nu}} (\alpha; t) & = \sum_{p_f} \Bigl( T^{(0)}_{\mu\nu}(p_f, p_i, \alpha ; t) T^{(0)*}_{\mu\nu}(p_f, p_i, \alpha ; t) + T^{(0)}_{\mu\nu}(p_f, p_i, \alpha ; t) \mathbb{E}\Bigl[T^{(2)*}_{\mu\nu}(p_f, p_i, \alpha ; t)\Bigr] \\
& \nonumber + \mathbb{E}\Bigl[T^{(2)}_{\mu\nu}(p_f, p_i, \alpha ; t)\Bigr] T^{(0)*}_{\mu\nu}(p_f, p_i, \alpha ; t) + \mathbb{E}\Bigl[ T^{(1)}_{\mu\nu}(p_f, p_i, \alpha ; t) T^{(1)*}_{\mu\nu}(p_f, p_i, \alpha ; t)\Bigr] \\
& \nonumber + \mathbb{E}\Bigl[T^{(4)}_{\mu\nu}(p_f, p_i, \alpha ; t)\Bigr] T^{(0)*}_{\mu\nu}(p_f, p_i, \alpha ; t) + T^{(0)}_{\mu\nu}(p_f, p_i, \alpha ; t) \mathbb{E}\Bigl[T^{(4)*}_{\mu\nu}(p_f, p_i, \alpha ; t)\Bigr] \\
& \nonumber + \mathbb{E}\Bigl[T^{(3)}_{\mu\nu}(p_f, p_i, \alpha ; t) T^{(1)*}_{\mu\nu}(p_f, p_i, \alpha ; t)\Bigr] + \mathbb{E}\Bigl[T^{(1)}_{\mu\nu}(p_f, p_i, \alpha ; t) T^{(3)*}_{\mu\nu}(p_f, p_i, \alpha ; t)\Bigr] \\
& \nonumber + \mathbb{E}\Bigl[T^{(2)}_{\mu\nu}(p_f, p_i, \alpha ; t) T^{(2)*}_{\mu\nu}(p_f, p_i, \alpha ; t)\Bigr]\Bigr)\;.
\end{align}
In the course of computation one has to take the average over the noise $\mathbb{E}$ leading to expressions that include integrals exemplarily of the form
\begin{align}
& \int\limits_0^t dt_1 \int\limits_0^{t_1} dt_2 \int\limits_0^{t_2} dt_3 \int\limits_0^{t_3} dt_4 \; \mathbb{E} [w(t_1) w(t_2)] \; \mathbb{E} [w(t_3) w(t_4)] = \int\limits_0^t dt_1 \int\limits_0^{t_1} dt_2 \int\limits_0^{t_2} dt_3 \int\limits_0^{t_3} dt_4 \delta(t_1 - t_2) \delta(t_3 - t_4) = \frac{1}{2} \theta^2(0) t^2,
\end{align}
where $\theta(0)$ is the Heaviside function at $x=0$. Calculating all the time integrals and adding the decay constants we obtain explicitly
\begin{align}
P_{K_{\mu=S/L}\rightarrow K_{\nu=S/L}} (\alpha; t) & = \delta_{\mu\nu}\Bigl[ 1 - \frac{\alpha}{2} \frac{\lambda m_{\mu}^2}{m_0^2} \Bigl( 2\theta(0) - 1\Bigr) t + \frac{3\alpha^2}{4} \frac{\lambda^2 m_{\mu}^4}{m_0^4} \Bigl(2\theta(0) (\theta(0) - 1 ) + \frac{1}{2} \Bigr) t^2 \Bigr]\cdot e^{-\Gamma_\mu t}\;.
\end{align}
The Heaviside function $\theta(x)$ at $x=0$ is not well defined, a value between zero and one~\cite{Berg1929, Kanwal1983} is generally possible. Taking three possibilities, $0$, $\frac{1}{2}$ or $1$, we obtain
\[
P_{K_{L/S}\rightarrow K_{L/S}} (\alpha; t) \;=\; e^{-\Gamma_{L/S} t}\cdot\begin{sqcases} 1 + \frac{1}{2} \frac{\lambda m_{L/S}^2}{m_0^2} \alpha t + \frac{3}{8} \frac{\lambda^2 m_{L/S}^4}{m_0^4} \alpha^2 t^2 \; &\mbox{if} \; \theta(0) = 0, \cr 1 \; &\mbox{if} \; \theta(0) = \frac{1}{2}, \cr 1 - \frac{1}{2} \frac{\lambda m_{L/S}^2}{m_0^2} \alpha t + \frac{3}{8} \frac{\lambda^2 m_{L/S}^4}{m_0^4} \alpha^2 t^2 \; &\mbox{if} \; \theta(0) = 1. \end{sqcases}
\]
Before we discuss the result let us also report the result for the strangeness-oscillations which is straightforward but much more cumbersome
\begin{eqnarray}
  P_{K^0 \rightarrow K^0/\bar{K}^0} &=& \frac{1}{4} \Biggl\{ e^{-\Gamma_L t}+e^{-\Gamma_S t} -  \frac{1}{2} \frac{\lambda}{m_0^2}\alpha t (m_L^2 e^{-\Gamma_L t}+ m_S^2 e^{-\Gamma_S t})\Bigl(2\theta(0)-1\Bigr)\nonumber\\
  &&\quad+ \frac{3}{4} \frac{\lambda^2}{m_0^4}\alpha^2 t^2 (m_L^4 e^{-\Gamma_L t}+ m_S^4e^{-\Gamma_S t})\Bigl(2\theta(0)(\theta(0)-1) + \frac{1}{2}\Bigr) \nonumber\\
 && \quad \pm 2\Biggl[1 - \frac{1}{2} \frac{\lambda}{m_0^2}\alpha t \Bigl( (m_L^2 + m_S^2)\theta(0) - m_L m_S\Bigr) + \frac{3}{8} \frac{\lambda^2}{m_0^4}\alpha^2 t^2 \Bigl( (m_L^4 + m_S^4)\theta^2(0) - 2m_L m_S (m_L^2 + m_S^2)\theta(0)\nonumber\\
 && \quad\quad + 2m_L^2 m_S^2 \Bigl(\theta^2(0) + \frac{1}{2} \Bigr) \Bigr) \Biggr]\cdot\cos\Bigl[(m_L - m_S)t\Bigr]\cdot e^{-\frac{\Gamma_L+\Gamma_S}{2} t}\Biggr\}\;.
\end{eqnarray}

The first choice $\theta(0)=0$ leads to probabilities greater than one if normalized to surviving mesons and the third choice $\theta(0)=1$ leads to probabilities generally decreasing in time, whereas the second choice conserves the particle number. The first option is certainly hard to motivate why during the dynamics particles are generated out of the noise field as well as the third case where one would lose some particles due to spontaneous collapse in position space. Therefore, for physical reasons the choice $\theta(0)=\frac{1}{2}$ is the most plausible one excluding all other choices in $\theta(0)\in[0,1]$, since in this case the particle number is conserved during the time evolution.

Heaviside functions defined by $\theta(0) = \frac{1}{2}$ can be approximated by continuous functions, for example $\theta(x) = \lim\limits_{k\rightarrow\infty} \frac{1}{1+e^{-2kx}}$. Note that this function value is also equivalent to the convention $\int_0^\infty\;\delta(t) dt\;=\; \frac{1}{2}$ for the Dirac-delta function $\delta(t)$ which is also often used for problems where the Laplace transformation helps solving and understanding the problem. Therefore, we may give this mathematical choice a proper physical meaning, i.e. corresponding to a particular physical property of the noise field. The noise in the QMUPL model then is considered to act continuously in time, not rapidly like a jump.

In summary, the obtained result for the QMUPL model shows that the physical properties of the noise field are dependent on the mathematical structure of the transition probabilities and a conservation of internal properties can be only obtained when the Heaviside function at zero is properly chosen. Due to the generality of the structure of collapse model our specific findings have to also hold for other collapse models. Consequently, our result paves the way to a new approach to investigate the physical properties of the noise field and to address an ontology.

The next step would be to compute the corresponding probabilities for the most popular collapse model, the mass-proportional Continuous Spontaneous Localization (CSL) model~\cite{GhirardiGrassiBenatti1995} and to investigate other systems with internal degrees of freedom.

\textbf{Acknowledgments:}
The authors gratefully thank the Austrian Science Fund (FWF-P26783) and the COST action MP1006.


\begin{thebibliography}{1}
\bibitem{GRW}
G.~C.~Ghirardi, A.~Rimini, T.~Weber, 
Phys. Rev. D 34 (1986) 470.

\bibitem{Catalina1}
C. Curceanu et al., 
Phys. Scr. 90 (2014) 028003.

\bibitem{Catalina2}
C. Curceanu, B. C. Hiesmayr and K. Piscicchia, 
J. Adv. Phys. 4 (2015) 263.

\bibitem{BassiDuerrHinrichs2011}
A.~Bassi, D.~D\"{u}rr, G.~Hinrichs, Phys. Rev. Lett. 111 (2011) 210401.

\bibitem{AdlerBassi2007}
S.~L.~Adler, A.~Bassi, J. Phys. A 40 (2007) 15083.
\bibitem{Diosi1989}
L.~Di\'{o}si, Phys. Rev. A 40 (1989) 1165.
\bibitem{Diosi1990}
L.~Di\'{o}si, Phys. Rev. A 42 (1990) 5086.
\bibitem{Berg1929}
E.~J.~Berg, \textit{Heaviside's operational calculus as applied to engineering and physics}, (1929) McGraw-Hill.

\bibitem{Kanwal1983}
R.~P.~Kanwal, \textit{Generalized Functions: Theory and Technique}, (1998) Birkhäuser.

\bibitem{GhirardiGrassiBenatti1995}
G.~C.~Ghirardi, R.~Grassi, F.~Benatti, Found. Phys. 25 (1995) 5.
\end{thebibliography}
\end{document}